\documentclass[12pt]{article}
\usepackage[dvips]{graphicx}
\usepackage[tight]{subfigure}
\usepackage{amssymb}

\def\be{\begin{equation}}
\def\ee{\end{equation}}
\def\bea{\begin{eqnarray}}
\def\eea{\end{eqnarray}}
\def\beaN{\begin{eqnarray*}}
\def\eeaN{\end{eqnarray*}}
\def\ed{\end{document}}
\def\bit{\begin{itemize}}
\def\eit{\end{itemize}}

\def\sig{\sigma}

\def\alf{\alpha}

\def\di{\partial}

\def\half{{\textstyle{1 \over 2}}}
\def\~{\tilde}
\def\lag{{{\cal L}}}
\def\m{\label}
\def\l{\left}
\def\r{\right}

\def\goto{\rightarrow}

\def\cG{\stackrel{\circ}{\Gamma}}

\def\cN{\stackrel{\circ}{\nabla}}

\def\cR{\stackrel{\circ}{R}}

\def\scJ{\stackrel{\bullet}{\cal J}}
\def\ccK{\stackrel{}{\cal K}}
\def\ccD{\stackrel{}{\cal D}}
\def\ccL{\stackrel{}{\cal L}}

\usepackage{authblk}

\usepackage[dvipsnames]{xcolor}
\usepackage{graphicx}

\begin{document}

\title{ \bf  Conserved quantities in STEGR and applications}
\author[1,2]{E. D. Emtsova\thanks{Electronic address: \texttt{ed.emcova@physics.msu.ru}}}
\author[1]{A. N. Petrov\thanks{Electronic address: \texttt{alex.petrov55@gmail.com}}}
\author[1,2]{A.V.Toporensky\thanks{Electronic address: \texttt{atopor@rambler.ru}}}
\affil[1]{Sternberg Astronomical institute, MV Lomonosov State university  \protect\\ Universitetskii pr., 13, Moscow, 119992,
Russia}
\affil[2]{Kazan Federal University, Kremlevskaya 18, Kazan, 420008, Russia}

\date{\small \today}
\maketitle

\begin{abstract}
  We derive conservation laws in Symmetric Teleparallel Equivalent of General Relativity (STEGR) with direct application of Noether’s theorem. This approach allows {us} to construct covariant conserved currents, corresponding superpotentials and invariant charges. A necessary component of our constructions is the concept of ``turning off'' gravity, introduced in the framework of STEGR to define the flat and torsionless connection. By calculating currents, one can obtain local characteristics of gravitational field like energy density. Surface integration of superpotentials gives charges which correspond to global quantities of the system like mass, momentum, etc. To test our results for the obtained currents and superpotentials, we calculate the energy density measured by freely falling observer in the simple solutions (Friedman universe, Schwartzchild black hole) and total mass of the Schwartzchild black hole. We find ambiguities in obtaining the connection, which explicitly affect the values of conserved quantities, and discuss possible solutions to this problem.
\end{abstract}

\section{Introduction}
General Relativity (GR) quite accurately describes modern astrophysical observations. However,  at the cosmological level, to satisfy the observations, one has to introduce phenomenologically additional structures: dark matter --- to explain  galaxy rotation curves, dark energy  --- to explain the accelerated  expansion of the universe. Theoretical explanation of dark matter and dark energy can be made by introducing various exotic models of elementary particles, or by different modified theories of gravity.

Most of modified gravity theories are based on the Riemannian curvature, but one can also modify general relativity basing on its equivalent formulations, such as Teleparallel Equivalent of General Relativity (TEGR), Symmetric Teleparallel Equivalent of General Relativity (STEGR) and other metric-affine theories ~\cite{BeltranJimenez:2019tjy}. Such modifications like $f(T)$, $f(Q)$, and others have attracted growing interest in recent years. From a theoretical point of view, torsion- and non-metricity-based theories are interesting because their field equations are second-order equations.

Torsion-based teleparallel theories of gravity including TEGR itself \cite{Aldrovandi_Pereira_2013}, and its various modifications, such as $f(T)$ gravity \cite{Cai:2015emx} and other models \cite{Bahamonde:2017wwk,Hohmann:2017duq} have been more developed and studied than non-metricity-based theories including STEGR with it's modification. However, the latter have been developing more and more intensively in recent years \cite{Lu:2021wif,Adak:2004uh,Adak:2005cd,Adak:2006rx,Adak:2008gd,Adak:2011ltj,Adak:2018vzk,Hohmann:2021fpr,BeltranJimenez:2017tkd,Jarv:2018bgs,Quiros:2022mut,BeltranJimenez:2021auj,Nakayama:2021rda,Dambrosio:2020wbi}. In the framework of symmetric teleparallel gravity, there are considered black hole solutions \cite{Wang:2021zaz,DAmbrosio:2021zpm,Mustafa:2021ykn,Zhao:2021zab}, cosmological problems \cite{Hohmann:2021ast,Sahoo:2021ctu,Kar:2021juu,Li:2021mdp,Bajardi:2020fxh,Zhao:2021zab,Gakis:2019rdd}, PPN formalism \cite{Flathmann:2020zyj} etc.  It often happens that problems which have arisen in  torsion-based teleparallel theories remain relevant in non-metricity-based theories.

In general relativity, there are problems in construction of  covariant conserved quantities: taking the Lagrangian of the first derivatives of dynamical variables only,
it is impossible to construct them without introducing additional structures. Indeed, then
 one obtains either coordinate non-covariant quantities in metric GR, or coordinate covariant but Lorentz non-covariant quantities in tetrad GR \cite{Petrov_KLT_2017}. In TEGR, this problem has been solved. For example, in papers \cite{Obukhov:2006ge,Obukhov_Rubilar_Pereira_2006}, using Noether theorem, fully covariant conserved quantities are constructed in the formalism of differential forms, which is not so popular in astrophysical and cosmological applications.
 %
 %
 In \cite{BeltranJimenez:2021kpj,Gomes:2022vrc} covariant Noether conserved quantities were constructed for general metric-affine gravity including TEGR and STEGR but their correspondence to physically expected values was not studied. In \cite{EPT19,EPT_2020}, first, fully covariant conserved quantities in TEGR were constructed in the tensorial formalism by direct application of the Noether theorem. And, second, it was shown that Noether's current describes the Einstein equivalence principle for the freely falling observers ``frozen'' into the Hubble flow in Friedmann–Lemaître–Robertson–Walker (FRLW) universe and (anti-)de Sitter space, and Noether's charge gives correct black hole mass for the Schwarzschild black hole.  In \cite{EKPT_2021,EKPT_2021a}, in TEGR, the Noether conserved quantities were studied in detail in different coordinate frames of the Schwarzschild solution, and the cases when they corresponded to physically meaningful ones where identified. In \cite{Emtsova:2021snt,Emtsova:2022ohe} it was shown that the Noether conserved quantities give correct mass and momentum for the moving Schwarzschild black hole.\footnote{Noether’s procedure can be useful not only for construction of conserved quantities. For example, in \cite{Paliathanasis:2014iva}, assuming a set of Noether’s symmetries in the framework of $f(T)$ theory, the authors find new solutions. The static spherically symmetric spacetime in $f(T)$ gravity is written in terms of the well-known Schwarzschild spacetime modified by a distortion function that depends on a characteristic radius. The obtained solutions are new ones and cannot be obtained by the usual methods.}

The purpose of this article is to begin a development of the research program carried out the framework of TEGR \cite{EPT19,EPT_2020,EKPT_2021,EKPT_2021a}  within the framework of STEGR. That is, using Noether's theorem, we derive conservation laws and corresponding conserved quantities in STEGR,  and then study their applications on the examples of the FRLW universe of the Schwarzschild black hole. The basis of the constructions in \cite{EPT19,EPT_2020,EKPT_2021,EKPT_2021a} was also the definition of generalized and universal ``turning off" gravity principle and generalized concept of ``gauges" in TEGR.
 Then the use of these constructions in applications was presented. Even within the framework of generalized and universal principles, problems and ambiguities have been found. These ambiguities have been studied and their interpretation have been given. Here, the ``turning off" gravity principle, similar to that in TEGR, is introduced and applied; the results of its application are discussed.

The paper is organized as follows. In the next section \ref{Elements} we introduce the main quantities of STEGR.
In section \ref{CL} we (i) describe the Noether pricedure, (ii) construct covariant conserved quantities with direct application of Noether’s theorem in STEGR, (iii) introduce the ``turning off'' gravity principle to obtain the flat and torsionless connection in STEGR.
In section \ref{App} we first find that the Noether current is in a correspondence with the Einstein equivalence principle for the freely falling observer in the  FRLW universe. Then we study if the Noether conserved quantities give physically expected values in different coordinates of the Schwarzschild solution: total black hole mass and a correspondance to the  Einstein equivalence principle for the freely falling observer.
In section \ref{Concl} we outline new results.

\section{Elements of STEGR}
\m{Elements}
\setcounter{equation}{0}
The Lagrangian in STEGR has the form \cite{BeltranJimenez:2019tjy}:
\begin{equation}\label{STlag}
  \ccL{} = \frac{\sqrt{-g}}{2 \kappa} g^{\mu\nu} (L^{\alpha} {}_{\beta \mu} L^{\beta} {}_{\nu \alpha} - L^{\alpha} {}_{\beta \alpha} L^{\beta} {}_{\mu\nu} ),
\end{equation}
where $\kappa=8 \pi$, and disformation is
\begin{equation}\label{defLofQ}
    L^{\alpha} {}_{\mu \nu}=\frac{1}{2} Q^{\alpha} {}_{\mu \nu} -\frac{1}{2} Q_{\mu} {}^{\alpha} {}_{\nu}-\frac{1}{2} Q_{\nu} {}^{\alpha} {}_{\mu}\,.
\end{equation}
Here, Greek indices are spacetime ones: $\alpha, \beta, \ldots = (0,1,2,3)$; Latin indices will denote only space ones: $i, j, \ldots = (1,2,3)$; the non-metricity $Q_{\alpha \mu \nu}$ is defined as follows:
\begin{equation}\label{defQ}
    Q_{\alpha \mu \nu} \equiv \nabla_\alpha g_{\mu \nu},
\end{equation}
where the covariant derivative $\nabla_\alpha$ is defined with the use of the connection $\Gamma^{\alpha} {}_{\mu \nu}$ which is symmetric in lower indexes and the corresponding torsion is zero: $T {}^\alpha {}_{\mu \nu} \equiv \Gamma {}^\alpha {}_{\mu \nu} - \Gamma {}^\alpha {}_{\nu\mu } = 0$. The curvature tensor of this connection is zero as well:
\begin{equation}\label{defRiem}
    R^{\alpha} {}_{\beta \mu \nu} (\Gamma) = \partial_{\mu} \Gamma^{\alpha} {}_{ \nu \beta} -  \partial_{\nu} \Gamma^{\alpha} {}_{\mu \beta} +  \Gamma^{\alpha} {}_{\mu \lambda}  \Gamma^{\lambda} {}_{\nu \beta} -  \Gamma^{\alpha} {}_{\nu \lambda}  \Gamma^{\lambda} {}_{\mu \beta} =0.
\end{equation}
One can easily verify that the decomposition of a general connection $\Gamma^{\beta} {}_{\mu \nu}$ into the Levi-Civita, contortion and the disformation terms, see \cite{BeltranJimenez:2019tjy}, reduces to
\begin{equation}\label{defL}
    L^{\beta} {}_{\mu \nu} \equiv \Gamma^{\beta} {}_{\mu \nu} - \cG{}^{\beta} {}_{\mu \nu},
\end{equation}
{where $L^{\beta} {}_{\mu \nu}$ and $\cG{}^{\beta} {}_{\mu \nu}$ are just the disformation and the Levi-Civita connenction, respectively.}

With the use of (\ref{defLofQ}) - (\ref{defL}) one can rewrite (\ref{STlag}) as
\begin{equation}\label{STlagtoHilbert}
      \ccL{} =   \ccL{}_{Hilb}+ \frac{\sqrt{-g} g^{\mu\nu}}{2 \kappa} R_{\mu\nu} +   \ccL{}' \equiv   - \frac{\sqrt{-g}}{2 \kappa} \cR + \frac{\sqrt{-g} g^{\mu\nu}}{2 \kappa} R_{\mu\nu} + \partial_\alpha  \ccD{}^\alpha,
\end{equation}
where the first term is the Hilbert Lagrangian and the third term $\ccL'$ is a total divergence:
\begin{equation}\label{defD}
      \ccL'{} =   - \frac{\sqrt{-g}}{2 \kappa}  \cN_\alpha (Q^\alpha-\hat{Q}^\alpha)=   \partial_\alpha \l(- \frac{1}{2 \kappa} \sqrt{-g} (Q^\alpha-\hat{Q}^\alpha)\r),
\end{equation}
and $Q_\alpha=g^{\mu\nu} Q_{\alpha \mu \nu}$, $\hat{Q}_\alpha=g^{\mu\nu} Q_{\mu \alpha \nu}$, $\cN_\alf$ is the Levi-Civita covariant derivative.

Let us discuss the second term $+ \frac{\sqrt{-g} g^{\mu\nu}}{2 \kappa} R_{\mu\nu}$ in (\ref{STlagtoHilbert}). If we vary (\ref{STlag}) directly we vary this term anyway, but only not explicitly. First, because $R^\alpha {}_{\mu \alpha \nu} = R_{\mu \nu}=0$ variation of this term with respect to $g^{\mu\nu}$ does not contribute into the field equations. Thus, in this sense, the term $+ \frac{\sqrt{-g} g^{\mu\nu}}{2 \kappa} R_{\mu\nu}$ is not sensible. On the other hand, the variation of (\ref{STlagtoHilbert}), the same (\ref{STlag}) with respect to $\Gamma^\alpha {}_{\mu\nu}$ gives the equation $\delta{\cal L}/\delta\Gamma^\alpha {}_{\mu\nu} =0$ that determines the teleparallel connection as  $\Gamma^\alpha {}_{\mu\nu} = \cG{} {}^\alpha {}_{\mu\nu}$. Due to (\ref{defL}) this restriction is very radical, and therefore impermissible. Thus, finally, under variation we consider the Lagrangian
\begin{equation}\label{Ls}
    \ccL{} =  - \frac{\sqrt{-g}}{2 \kappa} \cR  + \partial_\alf  \ccD{}^\alf
\end{equation}
instead of (\ref{STlagtoHilbert}), the same (\ref{STlag}), where
\begin{equation}\label{defD2}
    \ccD^{\alpha} \equiv  - \frac{\sqrt{-g}}{2 \kappa}  (Q^\alpha-\hat{Q}^\alpha)\,.
\end{equation}

\section{Conservation laws for STEGR Lagrangian}
\m{CL}
\setcounter{equation}{0}
Lagrangian (\ref{Ls}) is a scalar density, therefore one can apply directly a Noether theorem to obtain the related conserved current and superpotetntial.

\subsection{Noether conserved quantities in general}

The derivation of conserved quantities and related conservation laws for an arbitrary covariant theory of fields $\psi^A$ with an action
\be
S = \int dy^4x \lag(\psi^A;\psi^A{}_{,\alpha};\psi^A{}_{,\alpha\beta})\,.
\m{Sarbitrar}
\ee
follows from diffeomorphic invariance and can be found  in \cite{Mitskevich_1969,Lompay_Petrov_2013,Petrov_KLT_2017}.
Here, the fields $\psi^A$ are an arbitrary tensor density or a set of such densities, and $A$ is a collective index. Considering invariance of the theory (\ref{Sarbitrar}) with respect to diffeomorfisms, using the variations of fields $\psi^A$ in the form of  Lie derivatives of the fields $\psi^A$ \cite{Petrov_KLT_2017} one derives the conservation law for the Noether current $ {\cal J}{}^\alf(\xi)$:
\begin{equation}\label{CurrCL}
      \di_\alf{\cal J}^\alf(\xi)\equiv 0\,,
  \end{equation}
 from which follows that
\begin{equation}\label{CurrEqDSup}
 {\cal J}{}^\alf(\xi)=\partial_\beta  {\cal J}{}^{\alf\beta}(\xi),
\end{equation}
 where Noether superpotential $ {\cal J}{}^{\alf\beta}(\xi)$ is an antisymmetric tensor density.

Introducing the coefficients ${\cal U}_\sig{}^{\alf }$, ${\cal M}_\sig{}^{\alf\tau }$, ${\cal N}_\sig{}^{\alf\tau\beta}$ expressed through the Lagrangian derivatives in the form
 \bea {\cal U}_\sig{}^\alf & \equiv & \lag \delta^\alf_\sig +
 {{\delta \lag} \over {\delta \psi^B}} \l.\psi^B\r|^\alf_\sig -
 {{\delta \lag} \over {\delta  \psi^{B}{}_{,\alf}}} \di_\sig \psi^{B}  -  {{\di \lag} \over {\di \psi^{B}{}_{,\beta\alf}}} \di_{\beta\sig} \psi^{B}\, , \m{a-d8}\\
 {\cal M}_\sig{}^{\alf\tau} & \equiv &
 {{\delta \lag} \over {\delta  \psi^{B}{}_{,\alf}}}
 \l.\psi^{B}\r|^\tau_\sig -
 {{\di \lag} \over {\di  \psi^{B}{}_{,\tau\alf}}}
\di_\sig \psi^B +
 {{\di \lag} \over {\di \psi^{B}{}_{,\beta\alf}}}
 \di_\beta (\l.\psi^{B}\r|^\tau_\sig)\, ,
\m{a-d9}\\
{\cal N}_\sig{}^{\alf\tau\beta} & \equiv & {\frac12} \l[{{\di \lag} \over {\di  \psi^{B}{}_{,\beta\alf}}}
 \l.\psi^{B}\r|^\tau_\sig +
 {{\di \lag} \over {\di \psi^{B}{}_{,\tau\alf}}}
 \l.\psi^{B}\r|^\beta_\sig\r],
\m{a-d10}
 \eea
where the operation of the vertical line is defined as follows, see \cite{Petrov_KLT_2017}.
 We consider the Lie derivative of $\psi^A$ in the form
 $
{\pounds}_\xi \psi^A =-\xi^\alf \di_\alf \psi^A +
 {\l. \psi^A \r|}^\alf_\beta \di_\alf \xi^\beta\,,
\m{a-d4}
$
 note that we use the opposite signs with respect to the standard ones. The notation ${\l. \psi^A \r|}^\alf_\beta$ is defined by a concrete transformation properties of  $\psi^A$, for example, for the vector $\psi^A = \psi^\sigma$ one has ${\l. \psi^\sigma \r|}^\alf_\beta = \delta^\sigma_\beta \psi^\alpha$.
Then
the conserved current in (\ref{CurrCL}) has a form
\begin{equation}\label{defNetCurr}
    {\cal J}  {}^\alf(\xi) \equiv -({\cal U}_\sig{}^\alf\xi^\sig + {\cal M}_{\sig}{}^{\alf\tau}\di_\tau \xi^\sig r+ {\cal N}_{\sig}{}^{\alf\tau \beta}\di_{\tau \beta} \xi^\sig),
\end{equation}
and the superporential in (\ref{CurrEqDSup}) is
\begin{equation}\label{defNetSup}
    {\cal J}{}^{\alpha\beta} (\xi) = -\l( {\cal M}_\sig{}^{[\alf\beta]}\xi^\sig  - \frac{2}{3} \partial_\lambda {\cal N}_\sig{}^{[\alf \beta] \lambda}\xi^\sig + \frac{4}{3} {\cal N}_\sig{}^{[\alf \beta] \lambda} \partial_\lambda \xi^\sig  \r)\,.
\end{equation}

Finally, one defines a conserved quantity ${\cal P}(\xi)$ called as Noether charge:
\begin{equation}\label{ICQJa}
    {\cal P}(\xi) = \int_\Sigma d^3x {\cal J}^{0}(\xi)\,,
\end{equation}
 and using (\ref{CurrEqDSup}) and (\ref{defNetSup}) it effectively reduced to a surface integral
\begin{equation}\label{ICQJab}
   {\cal P}(\xi) =  \oint_{\di\Sigma} ds_i {\cal J}^{0i}(\xi)\,,
\end{equation}
where $\di\Sigma$ is a boundary of $\Sigma =: x^0 = t = {\rm const}$.

According to \cite{Petrov_KLT_2017} chapter 7, for every additional total divergence $div = \partial_\alpha {\cal D}^\alpha$, without considering its inner structure, it is possible to construct the corresponding additional parts of the current and the superpotential in the form:
\begin{equation}\label{divcurrsup}
 {\cal J}{}_{div}^\alf(\xi)=\partial_\beta  {\cal J}{}_{div}^{\alf\beta}(\xi)=\cN_\beta  {\cal J}{}_{div}^{\alf\beta}(\xi).
\end{equation}
Here, the additional  current ${\cal J}{}_{div}^{\alpha}$ and superpotential ${\cal J}{}_{div}^{\alpha\beta}$ for the divergence of ${\cal D}^\alpha$ are defined as:
\bea
\label{addcurr}
     {\cal J}{}_{div}^{\alpha}&=&\cN_\beta (-{\cal M}_{(div) \sigma} {}^{[\alpha\beta]} \xi^\sigma)= - {\cal U}_{(div) \sigma}{}^\alpha \xi^\sigma -{\cal M}_{(div) \sigma} {}^{[\alpha\beta]} \cN_\beta \xi^\sigma,\\
\label{addsup}
    {\cal J}{}_{div}^{\alpha\beta}&=&-{\cal M}_{(div) \sigma} {}^{[\alpha\beta]} \xi^\sigma,
\eea
where
\bea
\label{addm'}
 {\cal M}_{(div) \sigma}{}^{\alpha\beta}{}&=&2\delta_{\sigma}^{[\alpha} \ccD^{\beta]},\\
\label{addu'}
      {\cal U}_{(div) \sigma}{}^{\alpha}{}&=&2\cN_\beta(\delta_{\sigma}^{[\alpha} {\cal D}^{\beta]}) \equiv 2\partial_\beta(\delta_{\sigma}^{[\alpha} {\cal D}^{\beta]})\,.
\eea
The additional Noether charge can be defined in analogy to (\ref{ICQJa}), (\ref{ICQJab}):
\begin{equation}\label{ICQdiv}
    {\cal P}_{(div)}(\xi) = \int_\Sigma d^3x {\cal J}_{(div)}^{0}(\xi)=  \oint_{\di\Sigma} ds_i {\cal J}_{(div)}^{0i}(\xi).
\end{equation}
 So, despite the fact that adding total divergence to the Lagrangian does not affect the field equations, this divergence is explicitly added to the Noether current, superpotential, and charge.

\subsection{STEGR conserved quantities}
 Because all items in (\ref{Ls}) are scalar densities, it is natural to consider each item in (\ref{Ls}) separately under application of the Noether theorem. Considering the Hilbert Lagrangian, the current and superpotential are already derived in  \cite{Mitskevich_1969} and \cite{Petrov_KLT_2017}. Thus, the Noether current for the Hilbert Lagrangian is
\begin{equation}\label{defGRNetCurr}
    {\cal J}_{GR}^\alf(\xi) \equiv -( {\cal U}_\sig{}^\alf\xi^\sig + {\cal M}_{\sig}{}^{\alf\tau}\di_\tau \xi^\sig  + {\cal N}_{\sig}{}^{\alf\tau \beta}\di_{\tau \beta} \xi^\sig),
\end{equation}
where
\bea
\label{HilbN}
    {\cal N}_\sigma {}^{\alpha \tau \beta}& =& \frac{\sqrt{-g}}{4 \kappa} (2 g^{\tau \beta} \delta^\alpha_\sigma - g^{\alpha \beta} \delta^\tau_\sigma - g^{\alpha \tau} \delta^\beta_\sigma),\\
\label{HilbM}
   {\cal M}_\sigma {}^{\alpha \tau } &=&    \frac{\sqrt{-g}}{2 \kappa} (2 \cG{}^\alpha {}_{\sigma \omega} g^{\tau \omega} -  \cG{}^\omega {}_{\sigma \omega} g^{\alpha \tau} -    \cG{}^\tau {}_{\omega \epsilon} g^{\omega \epsilon} \delta^\alpha_\sigma),\\
     {\cal U}_\sigma {}^{\alpha} &=&    \frac{\sqrt{-g}}{2 \kappa} (g^{\alpha \lambda} g^{\omega \epsilon} - g^{\alpha \epsilon} g^{\omega \lambda} ) (g_{\lambda \sigma, \omega \epsilon} + \cG{}^\nu {}_{\omega \epsilon}  \cG{}_{\nu | \lambda \sigma} - \cG{}^\nu {}_{\omega \epsilon}   \cG{}_{\lambda | \nu  \sigma}  ) \nonumber \\
    &=&-  \frac{\sqrt{-g}}{\kappa} (G^\alf_\sigma+\frac{1}{2} \delta^\alf_\sigma R + \frac{1}{2} g^{\alpha \omega} \cG{} {}^{\rho} {}_{\rho (\omega,\sigma)}-\frac{1}{2} g^{\omega \epsilon} \cG{} {}^{\alpha} {}_{\omega \epsilon , \sigma})\label{HilbU}\,,
\eea
where the notation $\cG{}_{\alf | \beta  \gamma}$ means $\cG{}_{\alf | \beta  \gamma} = g_{\alf\rho}\cG^\rho{}_{ \beta  \gamma}$.

To represent the current and the superpotential in an explicitly covariant form we use the evident identities
\bea
\label{partialcov}
 \partial_\beta \xi^\sigma &\equiv &\cN_\beta \xi^\sigma - \cG{}^{\sigma} {}_{\lambda \beta} \xi^{\lambda},\\
     \partial_\tau \partial_\beta \xi^\sigma &\equiv &\cN_\tau  \cN_\beta \xi^\sigma  +  \l[\delta^\sigma_\lambda \cG{}^{\rho} {}_{\beta \tau}   -  \delta^\rho_\beta\cG{}^{\sigma} {}_{\lambda \beta} - \delta^\rho_\beta  \cG{}^{\sigma} {}_{\lambda \tau}\r] \cN_\rho  \xi^{\lambda}  \nonumber\\  & - & \l[\cG{}^{\sigma} {}_{\lambda \beta,\tau}  - \cG{}^{\sigma} {}_{\rho \beta} \cG{}^{\rho} {}_{\lambda \tau}\r]\xi^{\lambda}\label{partial2cov} .
\eea
Substituting (\ref{partialcov}) and  (\ref{partial2cov}) into (\ref{defGRNetCurr}), one rewrites the GR current in a covariant form:
\begin{equation}\label{defGRNetCurrCov}
    {\cal J}_{GR}^\alf(\xi) \equiv -( {\cal U*}_\sig{}^\alf\xi^\sig + {\cal M*}_{\sig}{}^{\alf\tau}\cN_\tau \xi^\sig + {\cal N*}_{\sig}{}^{\alf\tau \beta}\cN_{(\tau} \cN_{\beta)} \xi^\sig),
\end{equation}
with
\bea\label{HilbNcov}
    {\cal N*}_\sigma {}^{\alpha \tau \beta}& =& {\cal N}_\sigma {}^{\alpha \tau \beta},\\
    \label{HilbNcov}
   {\cal M*}_\sigma {}^{\alpha \tau } &=&   {\cal M}_\sigma {}^{\alpha \tau } - 2 {\cal N}_\lambda {}^{\alpha \tau \beta} \cG{}^{\lambda} {}_{\sigma \beta} + {\cal N}_\sigma {}^{\alpha \lambda \beta} \cG{}^{\tau} {}_{\beta \lambda } =0,\\
      {\cal U*}_\sigma {}^{\alpha} &=& {\cal U}_\sigma {}^{\alpha} - {\cal M}_\lambda {}^{\alpha \tau} \cG{}^{\lambda} {}_{\sigma \tau}  - {\cal N}_\lambda {}^{\alpha \tau \beta} \partial_\tau \cG{}^{\lambda} {}_{\sigma \beta}  + {\cal N}_\kappa {}^{\alpha \tau \beta} \cG{}^{\kappa} {}_{\lambda \beta} \cG{}^{\lambda} {}_{\sigma \tau}\nonumber\\ &=& - \frac{\sqrt{-g}}{4 \kappa} g^{\alpha \omega} \cR{}_{\omega \sigma}
      \label{HilbUcov}.
\eea

The superporential for the Hilbert Lagrangian after substituting (\ref{partialcov}), (\ref{HilbM}) and (\ref{HilbN}) into (\ref{defNetSup}) has the form
\begin{equation}\label{Komarsup}
     {\cal J}{}_{GR}^{\alpha\beta} =  \ccK{}^{\alpha\beta} = \frac{\sqrt{-g}}{\kappa} \cN{}^{[\alpha} \xi^{\beta]},
\end{equation}
that is well-known Komar's superpotential  \cite{Mitskevich_1969, Petrov_KLT_2017}.

Applying Noether theorem to the divergence $\ccL'{}$ in (\ref{defD}) with (\ref{defD2}), with the use of (\ref{addu'}) and \ref{addm'} one calculates additional Noether current (\ref{addcurr}), superpotential (\ref{addsup}) and charge (\ref{ICQdiv}).
Thus, the additional superpotential for $\ccL'{}$ in STEGR has an easy expression:
\begin{equation}\label{addsupstegr}
 {\cal J}{}_{div}^{\alpha\beta}= \frac{\sqrt{-g}}{\kappa} \delta_{\sigma}^{[\alpha}  (Q^{\beta]}-\hat{Q}^{\beta]}) \xi^\sigma.
\end{equation}
Integrating it by  (\ref{ICQdiv}) we get the additional Noether charge. Taking the divergence of (\ref{addsupstegr}) as (\ref{CurrEqDSup}) we get the additional Noether current same as (\ref{addcurr}).

Finally, the construction of the current and superpotential of the Lagrangian (\ref{Ls}) in STEGR is:
\begin{equation}\label{totalcurr}
    {\cal J}{}^{\alpha} = {\cal J}{}_{GR}^{\alpha} + {\cal J}{}_{div}^{\alpha},
\end{equation}
\begin{equation}\label{totalsup}
    {\cal J}{}^{\alpha\beta} = {\cal J}{}_{GR}^{\alpha\beta} + {\cal J}{}_{div}^{\alpha\beta}.
\end{equation}

By  construction, the current ${\cal J}{}^{\alf}(\xi)$ is a vector density, the superpotential ${\cal J}{}^{\alf\beta}(\xi)$ is an antisymmetric tensor density, and both are covariant. The Noether current is conserved:
\be
\di_\alf {\cal J}{}^\alf(\xi) = \cN_\alf {\cal J}{}^\alf(\xi) =0\,.
\m{DiffCL_2}
\ee
Integrating each part of (\ref{totalsup}) one gets
\begin{equation}\label{ICQtot}
    {\cal P}(\xi)={\cal P}_{GR}(\xi)+{\cal P}_{div}(\xi),
\end{equation}
where ${\cal P}(\xi)$ is (\ref{ICQJab}) and
\begin{equation}
    {\cal P}_{GR}(\xi) =  \oint_{\di\Sigma} ds_i {\cal J}_{GR}^{0i}(\xi),
\end{equation}
where ${\cal J}_{GR}^{0i}(\xi)$ is Komar superpotential (\ref{Komarsup});
\begin{equation}
     {\cal P}_{div}(\xi) =  \oint_{\di\Sigma} ds_i {\cal J}_{div}^{0i}(\xi),
\end{equation}
where ${\cal J}_{div}^{0i}(\xi)$ is (\ref{addsupstegr}).
\subsection{Defining the connection}
\m{turning_off}
To define the connection which is not a dynamical quantity we adapt for STEGR the ``turning off" gravity principle we developed in TEGR \cite{EPT19,EPT_2020,EKPT_2021,EKPT_2021a}.
Assuming that  $Q_{\alpha \mu \nu}$ and $L^{\alpha} {}_{\mu \nu}$ vanish in the absence of gravity and taking into account that $\cR{}^{\alpha} {}_{\beta \mu \nu}$ in GR  vanishes in the absence of gravity too we find the connection in STEGR as follows:

1) for known GR solution, construct related Riemann curvature tensor of the Levi-Civita connection:
\begin{equation}\label{Riemanntensor}
   \cR{}^\alpha{}_{\beta\mu\nu}=\di_\mu \cG{}^\alpha{}_{\beta\nu} - \di_\nu \cG{}^\alpha{}_{\beta\mu} + \cG{}^\alpha{}_{\kappa\mu}\cG{}^\kappa{}_{\beta\nu} - \cG{}^\alpha{}_{\kappa\nu}\cG{}^\kappa{}_{\beta\mu};
\end{equation}

2) to ``switch off” gravity solving the absent gravity equation $\stackrel{\circ}{R} {}^\alpha {}_{\beta \mu \nu}=0$ for parameters of the chosen GR solution;

3) When the parameters satisfying $\stackrel{\circ}{R} {}^\alpha {}_{\beta \mu \nu}=0$ are found, we take $\Gamma {}^{\alpha} {}_{ \mu \nu}=\stackrel{\circ}{\Gamma} {}^{\alpha} {}_{\mu \nu}$ for the found parameter values.


Torsion of the found connection should be zero because we take it from the Levi-Civita connection for some parameter values, and Levi-Civita connection is always symmetric. Curvature of the found connection should be zero too because we found it from the equation $\stackrel{\circ}{R} {}^\alpha {}_{\beta \gamma \delta}=0$.


\section{Applications}
\m{App}
\setcounter{equation}{0}
\subsection{Preliminary remarks}
Noether currents and superpotentials depend on the vector field $\xi$ that is not  fixed a priori, so we need  to find a way to determine it to obtain physically meaningful quantities.  A lot of ways to define $\xi$ exist, but which are physically meaningful?

  a) In the standard metric formulation of GR  the physically preferred choice is to use Killing fields of the reference geometry \cite{Chen:1998aw}.
  Time-like Killing vectors coincide with observers' proper vectors at the spatial infinity for isolated systems and we use them to obtain global characteristics of the object under consideration, which are total Noether charges in the studied approach. $\Sigma$ in (\ref{ICQJa}),(\ref{ICQtot}) is the whole infinite hypersurface of constant time in this case. It is expected that the result obtained by this way coincides with the mass of the object under consideration \cite{EPT19,EKPT_2021}.

  b) Obtaining local characteristics of the gravitational field we use the observer's proper vector (observer's 4-velocity).   It is expected that the components of Noether current (\ref{totalcurr}) give energy-momentum densities measured by the chosen observer \cite{EPT19,EKPT_2021}. The main experimental basis of general relativity (and other metric theories) is the weak equivalence principle. Of course, STERG as an equivalent representation of GR has to possess it. This means that a freely falling observer feels himself locally in Minkowski space. Thus, he does not measure any energetic characteristics. Calculating energy densities measured by freely falling observes, we check an applicability of suggested by us expressions for conserved quantities. In the case if the weak equivalence principle is not satisfied this means that the presumed interpretation of the conservation laws are invalid, and have to be changed
or specified. Note, for example, that the current explicitly depends on flat torsionless connection, see (\ref{divcurrsup}) for (\ref{addsupstegr}) with (\ref{defQ}), so we will try to obtain such flat connections which give zero current for the freely falling observer.  


We calculate energy desity measured by the freely falling observer in FRLW universe,  energy desity measured by the freely falling observer into the Schwarzschild black hole and Schwarzschild black hole total mass. For the  Schwarzschild solution we take static coordinates and general radially falling coordinates introduced by us in \cite{EKPT_2021,EKPT_2021a} which can be reduced to the Lemaitre coordinates in a particular case.

\subsection{FRLW universe}
We take the FRLW metric in the form:
\begin{equation}\label{metfrid}
    ds^2=-dt^2+a^2(t)\l(\frac{1}{1-kr^2}dr^2+r^2 \l(d\theta^2+ \sin ^2\theta  d\phi^2 \r)\r)
\end{equation}
where $k=+1$ for a positively curved space, $k=0$ for a flat space and $k=-1$ for a negatively curved space.

The non-zero components of Levi-Civita connection are:
\begin{equation}
    \begin{array}{cccc}
    \cG{}^{0} {}_{1 1} = \frac{a(t) a'(t)}{1-k r^2}; ~
\cG{}^{0} {}_{2 2} = r^2 a(t) a'(t); ~
\cG{}^{0} {}_{3 3} = r^2 a(t)  a'(t) \sin ^2\theta;\\
\cG{}^{1} {}_{0 1} =\cG{}^{1} {}_{1 0} = \cG{}^{2} {}_{0 2} =\cG{}^{2} {}_{2 0} = \cG{}^{3} {}_{0 3} =\cG{}^{3} {}_{3 0}   = \frac{a'(t)}{a(t)};\\
\cG{}^{1} {}_{1 1} = \frac{k r}{1-k r^2}; ~
\cG{}^{1} {}_{2 2} = r \left(k r^2-1\right); ~
\cG{}^{1} {}_{3 3} = r \sin ^2\theta \left(k r^2-1\right);\\
\cG{}^{2} {}_{1 2} = \cG{}^{2} {}_{2 1} =\cG{}^{3} {}_{1 3} =\cG{}^{3} {}_{3 1} = \frac{1}{r};\\
\cG{}^{2} {}_{3 3} =- \sin \theta  \cos \theta; ~
\cG{}^{3} {}_{2 3} = \cG{}^{3} {}_{3 2} = \cot \theta;\\
    \end{array}
\end{equation}
where numeration of coordinates is $(x^0,x^1,x^2,x^3)\equiv (t,r, \theta, \phi)$.

Non-zero components of Riemann tensor are proportional to:
\begin{equation}\label{fridriemtx}
\cR{}^0 {}_{i 0 i} \thicksim -\cR{}^0 {}_{i i0}  \thicksim \cR{}^i {}_{0 0 i} \thicksim -\cR{}^i {}_{0 i0} \thicksim \frac{\ddot{a}}{a},
\end{equation}
\begin{equation}\label{fridriemxx}
\cR{}^i {}_{j i j} \thicksim - \cR{}^i {}_{j ji} \thicksim  \frac{k+\dot{a}^2}{a^2}
\end{equation}
where $i,j = 1,2,3$. Equating these components to zero we got two equations which ``turn off'' gravity: $\dot{a}^2+k =0$ and $\ddot{a}=0$.
All solutions to the first equation satisfy the second equation, so only
\begin{equation}\label{fridvaceq}
\dot{a}^2+k =0
\end{equation}
makes sense. Note that  ``vacuum'' Friedmann equation
  \begin{equation}
      H^2=\rho_{curv}\qquad {\rm or}\qquad \left( \frac{\dot{a}}{a}\right)^2=-\frac{k}{a^2}\,,
      \m{H_vac}
  \end{equation}
is equivalent to  (\ref{fridvaceq}). The same condition was used to ``turn off'' gravity in TEGR  in FRLW universe \cite{EPT19,EPT_2020}.

Taking $a(t)= \sqrt{-k} t$ in  Levi-Civita connection we get the symmetric teleparallel connection:
\begin{equation}
    \begin{array}{cccc}
   \Gamma{}^{0} {}_{1 1} = \frac{k t}{k r^2-1};~
\Gamma{}^{0} {}_{2 2} = -k r^2 t;~
\Gamma{}^{0} {}_{3 3} = -k r^2 t \sin ^2\theta;\\
\Gamma{}^{1} {}_{0 1} = \Gamma{}^{1} {}_{1 0} = \Gamma{}^{2} {}_{0 2} =
\Gamma{}^{2} {}_{2 0} =\Gamma{}^{3} {}_{0 3} =\Gamma{}^{3} {}_{3 0} =\frac{1}{t};\\
 \Gamma{}^{1} {}_{1 1} = \frac{k r}{1-k r^2};~
\Gamma{}^{1} {}_{2 2} = r \left(k r^2-1\right);~
\Gamma{}^{1} {}_{3 3} = r \sin ^2\theta \left(k r^2-1\right);\\
 \Gamma{}^{2} {}_{1 2} =\Gamma{}^{2} {}_{2 1} =\Gamma{}^{3} {}_{1 3} = \Gamma{}^{3} {}_{3 1} = \frac{1}{r};\\
 \Gamma{}^{2} {}_{3 3} =- \sin \theta \cos \theta;~
\Gamma{}^{3} {}_{2 3} = \Gamma{}^{3} {}_{3 2} = \cot \theta;\\
  \end{array}
\end{equation}
where $(x^0,x^1,x^2,x^3)\equiv (t,r, \theta, \phi)$. Then we can obtain the non-metricity (\ref{defQ}) and disformation (\ref{defL}).

We consider a freely falling observer which is static in co-moving coordinates of the metric (\ref{metfrid}). Then components of his proper vector are:
\be
\xi^\sigma=(-1,0,0,0)\,.
\m{proper_FLRW}
\ee
We calculate the Noether current from the Noether  superpotential (\ref{totalsup}) using (\ref{CurrEqDSup}), where GR term is Komar superpotential (\ref{Komarsup}) and additional term is (\ref{addsupstegr}). The same result can be obtained calculating the Noether current  (\ref{totalcurr}) directly where GR term is obtained from (\ref{defGRNetCurr}) - (\ref{HilbU}) or (\ref{defGRNetCurrCov}),  (\ref{HilbNcov}), (\ref{HilbUcov}); and additional term is obtained from (\ref{addcurr}), (\ref{addm'}), (\ref{addu'}), (\ref{defD2}). The result for Noether current is:
 \be
 \scJ {}^{\alf } (\xi ) =(0,~0,~0,~0)\,.
 \m{J=0}
 \ee
   This means that the freely falling observer with the proper vector (\ref{proper_FLRW}) does not measure energy and momentum densities what corresponds to Einstein equivalence principle.

\subsection{Schwarzschild black hole}

The Schwarzschild metric in static coordinates has a form:
\begin{equation}
    ds^2=-fdt^2+f^{-1}dr^2+r^2 (d\theta^2 + \sin^2\theta d\phi^2),
    \label{BHmet}
\end{equation}
where
\begin{equation}\label{f(r)}
    f=f(r) =1-\frac{r_g}{r} = 1-\frac{2M}{r}
\end{equation}
and we keep $(t,r,\theta,\phi) = (x^0,x^1,x^2,x^3)$, respectively. The components of the timelike Killing vector in static coordinates are:
\begin{equation}\label{Killingtime}
\xi^\alpha = (-1,~0,~0,~0),
\end{equation}

After applying the coordinate transformations from the Schwarzschild static coordinates $(t,r,\theta, \phi)$ to Lemaitre freely falling coordinates $(\tau,\rho,\theta, \phi)$, preserving the numeration from $0$ to $3$,
\begin{equation}\label{SchtoLemTransf}
\begin{array}{cccc}
    d\rho =dt+ \frac{dr}{f\sqrt{1-f}}\,,  \\
d\tau =dt+\frac{dr}{f}\sqrt{1-f},
\end{array}
\end{equation}%
with $f(r)$ is defined in (\ref{f(r)}), the  Schwarzschild static metric (\ref{BHmet}) transforms to
\begin{equation} \label{metriclem}
ds^2=-d \tau^2+(1-f)d \rho^2+r^2 d\theta^2 + r^2 \sin^2 \theta d\phi^2 ,
\end{equation}
where
\begin{equation} \label{r_tau_rho}
r=r(\tau, \rho)=[\frac{3}{2} (\rho-\tau)]^{2/3} (2M)^{1/3}.
\end{equation}
Here, we are able to write $r$ as a function of $\rho$ and $\tau$ directly.

Solving the geodesic equation in general form for a radially freely falling observer towards the Schwarzschild black hole in the coordinates (\ref{BHmet}), one obtains the 4-velocity of radially freely falling observers with arbitrary energy parameter $e$. Written in Schwarzschild static coordinates expression of this 4-velocity is:
\begin{equation}
\label{velocityff}
\xi_{fall}^\alpha= \l(-\frac{e}{f},\sqrt{e^2-f},0,0\r)=\l(-\frac{e}{1-2M/r},\sqrt{e^2-(1-2M/r)},0,0\r).
\end{equation}
The parameter $e$ characterizes the initial state of the in-falling observer. In the case of $e>1$ the observer has a nonzero velocity directed to the black hole at spatial infinity $r\goto \infty$ and thus can reach each region of $r$. The observer with $e=1$ is connected to Lemaitre coordinates in (\ref{metriclem}), has a zero velocity at  the infinity $r\goto \infty$ and still able to achieve each point of space. In the case of $e<1$ the observer has zero velocity at some finite $r_0=r_0 (e)$ and cannot achieve the space region  $r>r_0 (e)$, and the proper vector of this observer does not exist at $r>r_0 (e)$.

Basing on such observers (\ref{velocityff}), one can construct the coordinate transformation from Schwarzschild static coordinates to general radially falling coordinates (which we call e-Lemaitre coordinates for brevity)  --- the freely falling observers' proper coordinates (in which the observers' 4-velocities would be $\xi_{fall}^\alf = (-1,0,0,0)$):
     \bea
          d \tau_e &=& edt +  \frac{\sqrt{e^2-1+\frac{2M}{r}}}{1-\frac{2M}{r}}   dr, \nonumber\\
          d \rho_e &=& dt +   \frac{e}{\left(1-\frac{2M}{r}\right) \sqrt{e^2-1+\frac{2M}{r}}} dr.
          \label{SchToFFcoord}
      \eea
Then, using (\ref{SchToFFcoord}), one transforms the Schwarzschild metric (\ref{BHmet}) to the form:
 \begin{equation}\label{FFmetric}
     ds^2=-d\tau_e^2 + \left(e^2-1+\frac{2M}{r}\right) d\rho_e^2  + r^2\left( d\theta^2 + \sin^2 \theta d\phi^2 \right)\,,
 \end{equation}
 where $r = r(\tau_e,\rho_e)$.

 These coordinates  contain the classical Lemaitre coordinates as a particular case ($e=1$), being in its turn
 a particular example of more general construction \cite{Bronnikov, Zaslavskii}. In the e-Lemaitre metric
 the dependence of
  $r$ as a function of $\rho_e$ and $\tau_e$ is not so simple as in the classical Lemaitre case (\ref{r_tau_rho}), and
  we will try to avoid to use this function directly in what follows.

 \subsection{Calculations in static case}
 Non-zero Levi-Civita connection components calculated for the metric (\ref{BHmet}) are:
 \begin{equation}\label{LCconnstat}
 \begin{array}{cccc}
    \cG{}^{0} {}_{0 1} =\cG{}^{1} {}_{1 1} = \frac{{r_g}}{2r^2}\frac{1}{1- {r_g}/r};~\cG{}^{1} {}_{0 0} = \frac{{r_g}}{2r^2} (1-\frac{r_g}{r});~~\cG{}^{2} {}_{1 2} = \cG{}^{3} {}_{1 3} = \frac{1}{r};\\

    \cG{}^{1} {}_{2 2} = -r\l(1-\frac{r_g}{r}\r);~\cG{}^{1} {}_{3 3} = -r\l(1-\frac{r_g}{r}\r)\sin^2\theta;\\

\cG{}^{2} {}_{3 3} = - \sin \theta \cos \theta ;~~\cG{}^{3} {}_{2 3} = \cot \theta.\\
 \end{array}
 \end{equation}
 Then {only non-zero} components of the Komar superpotantial of GR (\ref{Komarsup}) for Killing vector (\ref{Killingtime}) are
\begin{equation}\label{Komarkill}
\ccK{}^{01}  = - \ccK{}^{10}  =  \frac{r_g} {16 \pi }\sin \theta .
\end{equation}

``Turning off'' gravity in (\ref{LCconnstat}) by $r_g \goto 0$ we get:
\begin{equation}\label{STconnstat}
\cG{}^{2} {}_{1 2} = \cG{}^{3} {}_{1 3} = {1}/{r};~
\cG{}^{1} {}_{2 2} = -r;~\cG{}^{1} {}_{3 3} = -r\sin ^2\theta;~\cG{}^{2} {}_{3 3} = - \sin \theta \cos \theta;~\cG{}^{3} {}_{2 3} = \cot \theta.
\end{equation}
{Recently, see (33) in \cite{Lin_Zhai_2021}, the same teleparallel connection components have been derived {in $f(Q)$ theories}  for the static spherical solutions in the standard coordinates by ``turning off'' gravity.}
For (\ref{LCconnstat}) and (\ref{STconnstat}) for 01-component and 10-component of (\ref{addsup}) with (\ref{addm'}) and (\ref{defD2}) and the  Killing vector (\ref{Killingtime}) one has
 \begin{equation}\label{currdivkill}
       {\cal J}_{div}^{01} (\xi_{kill}) = - {\cal J}_{div}^{10} (\xi_{kill}) = \frac{r_g} {16 \pi }\sin \theta.
 \end{equation}
Summing (\ref{Komarkill}) and (\ref{currdivkill}) by (\ref{totalsup}) and providing the calculation of charge (\ref{ICQJab}), (\ref{ICQtot}), we obtain for the total mass:
\begin{equation}
    E=\lim_{r\rightarrow\infty} \oint_{\di\Sigma}  {\cal J}{}^{01}(\xi_{kill})d \theta d \phi =\frac{M}{2}+\frac{M}{2}=M,
\end{equation}
 that is the {\em acceptable result}.

Let us consider a freely falling observers. Then the 01-component and 10-component of the Komar superpotantial of GR (\ref{Komarsup})  with the freely falling observer's vector (\ref{velocityff}) are zero
\begin{equation}
\ccK{}^{01} = - \ccK{}^{10} = 0
\end{equation}
without dependence on $e$ {the same as other components $\ccK{}^{\alf\beta}=0$. As a result,} the related current by (\ref{CurrEqDSup}) is zero too.

For the additional superpotential (\ref{addsup}) for (\ref{LCconnstat}) and (\ref{STconnstat})  with the freely falling observer's vector (\ref{velocityff})  one has
 \begin{equation}
      {\cal J}_{div}^{01} (\xi_{fall}) = -  {\cal J}_{div}^{10} (\xi_{fall}) =\frac{e {r_g}\sin\theta}{16 \pi(1  - {r_g}/r)}
 \end{equation}
{with other components of ${\cal J}_{div}^{\alf\beta}(\xi_{fall}) = 0$.} Then, by (\ref{divcurrsup}) we obtain the current:
 \begin{equation}
    {\cal J}^\alpha_{div} (\xi_{fall})=  \left(-\frac{e {r_g}^2\sin\theta}{16 \pi r^2(1  - {r_g}/r)^2},0,0,0\right).
 \end{equation}
 Thus, the total current components are non-zero that {\em is not in correspondence with the equivalence principle}.

 Principally, the results of this subsection repeat the results obtained in \cite{EKPT_2021},\cite{EKPT_2021a} in the {\em static gauge}.

\subsection{The e-Lemaitre case.}

We calculate the Levi-Civita connection components for the metric (\ref{FFmetric}):
\begin{equation}
    \begin{array}{cccc}
 \cG{}^{0} {}_{1 1} = \frac{{r_g}}{2r^2}  \sqrt{e^2-1+{r_g}/r};~\cG{}^{0} {}_{2 2} = -r\sqrt{e^2-1+{r_g}/r};\\
 \cG{}^{0} {}_{3 3} = -r \sin ^2\theta \sqrt{e^2-1+{r_g}/r};\\

\cG{}^{1} {}_{0 1} = \frac{{r_g}}{2r^2} \l({\sqrt{e^2-1+{r_g}/r}}\r)^{-1};~\cG{}^{1} {}_{1 1} = - \frac{{er_g}}{2r^2} \l({\sqrt{e^2-1+{r_g}/r}}\r)^{-1};\\
 \cG{}^{1} {}_{2 2} = - {er}\l({\sqrt{e^2-1+{r_g}/r}}\r)^{-1};~\cG{}^{1} {}_{3 3} = - {er} \sin^2\theta\l({\sqrt{e^2-1+{r_g}/r}}\r)^{-1};\\

\cG{}^{2} {}_{0 2} = \cG{}^{3} {}_{0 3} =- \frac{{1}}{r}  \sqrt{e^2-1+{r_g}/r};~\cG{}^{2} {}_{1 2} = \cG{}^{3} {}_{1 3} = \frac{{e}}{r}  \sqrt{e^2-1+{r_g}/r};\\

\cG{}^{2} {}_{3 3} = - \sin \theta \cos \theta;~\cG{}^{3} {}_{2 3} = \cot\theta.\\
    \end{array}
\end{equation}
``Turning off'' gravity by $r_g \goto 0$, where $r$ is an \textit{independent on $r_g$ coordinate}  (function) we get the teleparallel connection components with the $e$-case indices in coordinates $(\tau_e, \rho_e, \theta, \phi)$:
\begin{equation}
    \begin{array}{cccc}\label{1st_switch}
\Gamma{}^{0} {}_{2 2} = -r\sqrt{e^2-1};~ \Gamma{}^{0} {}_{3 3} = -r \sin ^2\theta \sqrt{e^2-1};\\

 \Gamma{}^{1} {}_{2 2} = - {er}\l(\sqrt{e^2-1}\r)^{-1};~\Gamma{}^{1} {}_{3 3} = - {er} \sin^2\theta\l(\sqrt{e^2-1}\r)^{-1};\\

\Gamma{}^{2} {}_{0 2} = \cG{}^{3} {}_{0 3} =- \frac{{1}}{r}  \sqrt{e^2-1};\Gamma{}^{2} {}_{1 2} = \Gamma{}^{3} {}_{1 3} = \frac{{e}}{r}  \sqrt{e^2-1};\\

\Gamma{}^{2} {}_{3 3} = - \sin \theta \cos \theta;~\Gamma{}^{3} {}_{2 3} = \cot\theta.\\
    \end{array}
\end{equation}
Then transforming this connection components to the Schwarzschild coordinates by the usual way we get for the further calculations in convenient Schwarzschild coordinates:
\begin{equation}
    \begin{array}{cccc}
 \Gamma{}^{0} {}_{1 1} = -\frac{e {r_g}}{r^2(r-{r_g/r})^2 }\frac{1}{\sqrt{e^2-1+{r_g}/r}};~
\Gamma{}^{0} {}_{2 2} = \frac{e {r_g}}{1-{r_g}/r}\frac{1}{\sqrt{e^2-1} } ;\\

\Gamma{}^{0} {}_{3 3} = \frac{e {r_g}\sin ^2\theta}{1-{r_g}/r}\frac{1}{\sqrt{e^2-1} };~
\Gamma{}^{1} {}_{1 1} = \frac{{r_g}}{2 r^2} \frac{1}{e^2-1+{r_g}/r};\\

\Gamma{}^{1} {}_{2 2} = -r\sqrt{\frac{e^2-1+{r_g}/r}{e^2-1}};~\Gamma{}^{1} {}_{3 3} =-r\sqrt{\frac{e^2-1+{r_g}/r}{e^2-1}}\sin^2\theta; \\

\Gamma{}^{2} {}_{1 2} = \Gamma{}^{3} {}_{1 3} = \frac{1}{r}\sqrt{\frac{e^2-1}{e^2-1+{r_g}/r}};~\Gamma{}^{2} {}_{3 3} = - \sin \theta \cos \theta;~\Gamma{}^{3} {}_{2 3} = \cot\theta.\\

    \end{array}
\end{equation}
It turns out that this way of the ``switching-off'' is not correct because Riemannian tensor is not zero for this connection and thus we cannot use it anymore.

We think that the reason of this unacceptable result is that we did not ``turn off'' gravity totally,  i.e. we did not provide $r_g \to 0$ inside function $r(\tau_e,\rho_e)$ because really it is not independent function, since we made a coordinate transformation to  freely falling coordinates $(\tau_e,\rho_e,\theta,\phi)$. Calculating Riemann tensor (\ref{Riemanntensor}) of the connection (\ref{1st_switch}) directly one can use partial derivatives taken from (\ref{SchToFFcoord}) and see that  Riemann tensor components depend on  $r_g$.

To provide a correct ``switching off'' gravity setting $r_g=0$ we introduce the operation $r(\tau_e,\rho_e) \stackrel{r_g\goto 0}{\longrightarrow} \~r(\tau_e,\rho_e)$ where $\~r(\tau_e,\rho_e)$ totally does not depend on $r_g$.
To define the function  $\~r(\tau_e,\rho_e)$ we do the following. From the coordinate transformations (\ref{SchToFFcoord}) we have
 \be
          d (\tau_e -e \rho_e ) =  \left(  \frac{\sqrt{e^2-1+\frac{r_g}{r}}}{1-\frac{r_g}{r}} -\frac{e^2}{\left(1-\frac{r_g}{r}\right) \sqrt{e^2-1+\frac{r_g}{r}}} \right)dr  =
          -\frac{dr}{\sqrt{e^2-1+\frac{r_g}{r}}}.
      \ee
After integration one has
 \bea
 \tau_e -e \rho_e  &=& -\frac{\sqrt{r\l(r(e^2 - 1)+{r_g}\r)}}{e^2-1}\nonumber\\
 &+& \frac{r_g}{(e^2-1)^{3/2}}\l\{\log\l[\sqrt{r(e^2-1)+r_g} + \sqrt{r(e^2-1)}\r]+C\r\},
 \m{int}
 \end{eqnarray}
 where $C$ is the integration constant.  Let us ``switch off'' gravity in this relation
by $r_g=0$, as usual. Then, we obtain $\~r$ without dependence on $r_g$ as

\begin{equation}
    \tau_e -e \rho_e = -\frac{\~r}{\sqrt{e^2-1}}.
\end{equation}
So, comparing last two equations we get
 \be\label{r-to-r0}
{\~r} = \frac{\sqrt{r\l(r(e^2 - 1)+{r_g}\r)}}{\sqrt{e^2-1}} -
 \frac{r_g}{e^2-1}\l\{\log\l[\sqrt{r(e^2-1)+r_g} + \sqrt{r(e^2-1)}\r]+C\r\}.
 \ee

 Substituting the last expression into (\ref{1st_switch}), we are ``turning off'' gravity in external way and in internal way inside $r$ that goes to $\~r$.
Under this paradigma, the teleparallel connection in e-falling coordinates is
\begin{equation}\label{econnshort}
    \begin{array}{cccc}
  \Gamma{}^{0} {}_{2 2} = -\~r\sqrt{e^2-1};~\Gamma{}^{0} {}_{3 3} = -\~r\sqrt{e^2-1}\sin ^2\theta ;~\Gamma{}^{1} {}_{2 2} = -\frac{e \~r}{\sqrt{e^2-1}};\\

\Gamma{}^{1} {}_{3 3} = -\frac{e \~r \sin ^2\theta }{\sqrt{e^2-1}};~\Gamma{}^{2} {}_{0 2} = \Gamma{}^{3} {}_{0 3} =-\frac{\sqrt{e^2-1}}{\~r};~\\

\Gamma{}^{2} {}_{1 2} = \Gamma{}^{3} {}_{1 3} = \frac{e \sqrt{e^2-1}}{\~r};~\Gamma{}^{2} {}_{3 3} = -\sin \theta \cos \theta;~\Gamma{}^{3} {}_{2 3} = \cot \theta.\\

    \end{array}
\end{equation}
So, the connection in e-falling coordinates is
\begin{equation}
    \begin{array}{cccc}
\Gamma{}^{0} {}_{2 2} = -\sqrt{r \left(r\left(e^2-1\right)+{r_g}\right)}+ \frac{{r_g} \log \l(\sqrt{r\left(e^2-1\right)+{r_g}}+ \sqrt{r\left(e^2-1\right)}\r)+C}{\sqrt{e^2-1}};\\

\Gamma{}^{0} {}_{3 3} = \l[-\sqrt{r \left(r\left(e^2-1\right) +{r_g}\right)}+\frac{ {r_g}\log \left(\sqrt{r\left(e^2-1\right)+{r_g}}+ \sqrt{r\left(e^2-1\right)}\right)+C}{\sqrt{e^2-1}}\r]\sin^2\theta;\\

\Gamma{}^{1} {}_{2 2} = e\l[-\frac{\sqrt{r \left(r\left(e^2-1\right) +{r_g}\right)}}{e^2-1}+\frac{ {r_g}\log \left(\sqrt{r\left(e^2-1\right)+{r_g}}+ \sqrt{r\left(e^2-1\right)}\right)+C}{{e^2-1}}\r];\\

\Gamma{}^{1} {}_{3 3} = e\l[-\frac{\sqrt{r \left(r\left(e^2-1\right) +{r_g}\right)}}{e^2-1}+\frac{ {r_g}\log \left(\sqrt{r\left(e^2-1\right)+{r_g}}+ \sqrt{r\left(e^2-1\right)}\right)+C}{{e^2-1}}\r]\sin^2\theta;\\

\Gamma{}^{2} {}_{0 2} = \Gamma{}^{3} {}_{0 3} =\frac{e^2-1}{- \sqrt{ r \left(\left(e^2-1\right) r+{r_g}\right)}+ {r_g} \log \left(\sqrt{ \left(e^2-1\right) r+{r_g}}+ \sqrt{r\left(e^2-1\right)}\right)+C};\\

\Gamma{}^{2} {}_{1 2} = \Gamma{}^{3} {}_{1 3} =\frac{ e\left(e^2-1\right)}{\sqrt{ r \left(\left(e^2-1\right) r+{r_g}\right)}- {r_g} \log \left(\sqrt{ \left(e^2-1\right) r+{r_g}}+ \sqrt{r\left(e^2-1\right)}\right)-C};\\

\Gamma{}^{2} {}_{3 3} = - \sin\theta\cos\theta;~~\Gamma{}^{3} {}_{2 3} = \cot\theta.\\

 \end{array}
\end{equation}

 Connection after transformation to static coordinates takes the form
 \begin{equation}\label{e_L_connection}
    \begin{array}{cccc}
\Gamma{}^{0} {}_{1 1} = -\frac{e {r_g}\sqrt{r}}{(r-{r_g})^2 \sqrt{r(e^2-1)+{r_g}}};~~\Gamma{}^{1} {}_{1 1} = \frac{{r_g}}{2 r \left(\left(e^2-1\right) r+{r_g}\right)};\\

\Gamma{}^{0} {}_{2 2} = {e {r_g}} \l[\frac{\sqrt{ r \left(r\left(e^2-1\right) +{r_g}\right)}}{\left(e^2-1\right)(r-{r_g})} - \frac{{r_g} \log \left(\sqrt{\left(r\left(e^2-1\right) +{r_g}\right)}+ \sqrt{r\left(e^2-1\right)}\right)-C}{\left(e^2-1\right)(r-{r_g})} \r];\\

\Gamma{}^{0} {}_{3 3} = e {r_g} \l[\frac{\sqrt{ r \left(\left(e^2-1\right) r+{r_g}\right)}}{\left(e^2-1\right)(r-{r_g})} - \frac{{r_g} \log \left(\sqrt{\left(r\left(e^2-1\right) +{r_g}\right)}+ \sqrt{r\left(e^2-1\right)}\right)-C}{\left(e^2-1\right) (r-{r_g})}\r]\sin^2\theta;\\

\Gamma{}^{1} {}_{2 2} = - \sqrt{e^2+\frac{{r_g}}{r}-1}\l[\frac{\sqrt{ r \left(\left(e^2-1\right) r+{r_g}\right)}}{\left(e^2-1\right)} - \frac{{r_g} \log \left(\sqrt{\left(r\left(e^2-1\right) +{r_g}\right)}+ \sqrt{r\left(e^2-1\right)}\right)-C}{\left(e^2-1\right) }\r];\\

\Gamma{}^{1} {}_{3 3} = - \sqrt{e^2+\frac{{r_g}}{r}-1}\l[\frac{\sqrt{ r \left(\left(e^2-1\right) r+{r_g}\right)}}{\left(e^2-1\right)} - \frac{{r_g} \log \left(\sqrt{\left(r\left(e^2-1\right) +{r_g}\right)}+ \sqrt{r\left(e^2-1\right)}\right)-C}{\left(e^2-1\right) }\r]\sin^2\theta;\\

\Gamma{}^{2} {}_{1 2} = \Gamma{}^{3} {}_{1 3} =\frac{ e^2-1}{\sqrt{e^2+\frac{{r_g}}{r}-1}\l(\sqrt{ r \left(\left(e^2-1\right) r+{r_g}\right)}- {r_g} \log \left(\sqrt{ \left(e^2-1\right) r+{r_g}}+ \sqrt{r\left(e^2-1\right)}\right)-C\r)};\\

\Gamma{}^{2} {}_{3 3} = - \sin\theta\cos\theta;~~\Gamma{}^{3} {}_{2 3} = \cot\theta.\\

  \end{array}
\end{equation}
 Calculating the Riemannian tensor components for this connection we find that all of them
present fractions where the numerators are equal to zero identically. This means that
despite the denominators can vanish for some particular $r= r(C)$, the curvature equals zero
for all permissible $r$. So that, the connection (\ref{e_L_connection}),
despite it is singular at the point when the denominator vanishes, is a teleparallel connection.


{For the sake of simplicity and certainty let us choose $C = -\half r_g\log\sqrt{e^2-1}$. Then,} the only non-zero components of the additional divergence superpotential with Killing vector (\ref{Killingtime}) are
\begin{equation}
\begin{array}{cccc}
 {\cal J}^{01}_{div}(\xi_{kill}) = - {\cal J}^{10}_{div}(\xi_{kill}) =  \left({r_g} r^2 \sin\theta\left( 2r\left(e^2-1\right)^3 r+e^2 \left(3 e^2-4\right) {r_g}+{r_g}\right)\right.  \\
+{r_g} \l[2\log \left(\sqrt{ \left(e^2-1\right) r+{r_g}}+ \sqrt{r\left(e^2-1\right)}\right)- \log\sqrt{e^2-1} \r]*  \\ \left.
*   \left(\left(3 e^2+1\right) \sqrt{r(e^2 - 1)(r(e^2-1)+r_g)} \right.\right. \\ \left.\left.
 +   \left(\left(e^2-1\right) r+{r_g}\right) \l[2\log \left(\sqrt{ \left(e^2-1\right) r+{r_g}}+ \sqrt{r\left(e^2-1\right)}\right)- \log\sqrt{e^2-1} \r] \right)\right)/  \\
(16 \pi  \left(e^2-1\right) r^3 \left(2 \left(e^2-1\right) \left(\left(e^2-1\right) r+{r_g}\right) + (r_g/r)\sqrt{r(e^2 - 1)(r(e^2-1)+r_g)}~~*\right. \\ \left.
 *\log
 \l[2\log \left(\sqrt{ \left(e^2-1\right) r+{r_g}}+ \sqrt{r\left(e^2-1\right)}\right)- \log\sqrt{e^2-1} \r] \right) ). \\
\end{array}
\end{equation}
This gives the additional $+M/2$ to Komar  $M/2$ to obtain the acceptable value $M$. This result is valid for all $C$ introduced in (\ref{int}) as well.

The only non zero components of the additional divergence superpotential with a free falling observer's vector (\ref{velocityff}) are
\begin{equation}
\begin{array}{cccc}
 {\cal J}^{01}_{div}(\xi_{fall}) = - {\cal J}^{10}_{div}(\xi_{fall}) =    \left(e {r_g}^2r^2 \sin\theta \left(\left(e^2-1\right) r+{r_g}\right)*\right.  \\
* \l[2\log \left(\sqrt{ \left(e^2-1\right) r+{r_g}}+ \sqrt{r\left(e^2-1\right)}\right)- \log\sqrt{e^2-1} \r]^2  \\ 
+   4\sqrt{r(e^2 - 1)(r(e^2-1)+r_g)} * \\ \left.
 * \l[2\log \left(\sqrt{ \left(e^2-1\right) r+{r_g}}+ \sqrt{r\left(e^2-1\right)}\right)- \log\sqrt{e^2-1} \r] +4 \left(e^2-1\right)\right)/  \\
(16 \pi  \left(e^2-1\right) r^3 \left(2 \left(e^2-1\right) \left(\left(e^2-1\right) r+{r_g}\right) + (r_g/r)\sqrt{r(e^2 - 1)(r(e^2-1)+r_g)}~~*\right. \\ \left.
* \log
 \l[2\log \left(\sqrt{ \left(e^2-1\right) r+{r_g}}+ \sqrt{r\left(e^2-1\right)}\right)- \log\sqrt{e^2-1} \r] \right) ). \\
\end{array}
\end{equation}
So, the 0-component of current is not zero. We see that unlike the TEGR, we got the correct black hole mass in static case like in TEGR, but, unlike the TEGR, we couldn't get Einstein equivalence principle for the freely falling observer not only in static case but in the freely falling case too.

Note, that some components in (\ref{econnshort}) tend to infinity when $e \goto +1$.
Thus, substituting this connection into the expressions for current and superpotential which explicitly depend on the flat torsionless connection, see (\ref{divcurrsup}) for (\ref{addsupstegr}) with (\ref{defQ}), we get infinite values too, although the flat connection components are not observable. ``Turning off'' gravity in Lemaitre  coordinates directly leads also to infinite connection  or non-zero Riemann tensor  values, although it has to be zero for the flat connection. So, ``turning off'' gravity  principle might be unsuccessful in certain cases. Here, for example, it is unsuccessful with making the use of the Lemaitre coordinates for the Schwarzschild solution. In this particular case, the reason is in that the transformation (\ref{SchtoLemTransf}) is singular with respect to $r_g$. This is a problem of calculation methods only, and has to be resolved by a related regularization procedure.
This suggests also that we could probe other ways of obtaining the connection in STEGR.

\section{Conclusions}
\m{Concl}

Applying Noether theorem we have constructed in tensorial form covariant conserved quantities in STEGR: currents, superpotentials and charges. Because STEGR Lagrangian can be written as the Hilbert Lagrangian plus the total divergence, these quantities consist of two components: GR term plus the divergent term, see (\ref{totalcurr}), (\ref{totalsup}), (\ref{ICQtot}).  Each term is expressed in evidently covariant form: (\ref{Komarsup}) and (\ref{addsupstegr}) for the Noether superpotential, covariant derivatives of (\ref{Komarsup}) and (\ref{addsupstegr})  give the Noether current, integration (\ref{ICQJab}) of (\ref{Komarsup}) and (\ref{addsupstegr}) gives the Noether charge. All these expressions are covariant under coordinate transformations: GR parts are covariant due to the covariance of the Komar superpotential and covariant derivatives of it; the superpoential (\ref{addsupstegr}) related to the divergent term  and the corresponding current $ {\cal J}{}_{div}^\alf(\xi)=\partial_\beta  {\cal J}{}_{div}^{\alf\beta}(\xi)$ are proportional to non-metricity (\ref{defQ}); the same, to the disformation tensor (\ref{defL})  (which are coordinate-covariant due to introducing of flat torsionless connection) and their covariant derivatives. However, the flat torsionless connection in STEGR is undetermined --- it cannot be obtained by the field equations. To solve this problem we find the connection using the ``turning off'' gravity principle, the idea for which was taken from TEGR ``turning off'' gravity principle, and the method itself was then adapted for STEGR. Usually, authors aim to find a ``coincident" gauge with zero connection corresponding to particular coordinates \cite{Bahamonde:2022zgj,Wang:2021zaz}. In our case of covariant expressions, we can obtain the necessary teleparallel connection for each coordinates, and then make coordinate transformation to ``coincident" gauge.

To test the applications for the new Noether  conserved quantities we consider FRLW and Schwartzchild solutions. For the Friedmann solution we obtained zero Noether current for the freely falling observers ``frozen" into the Hubble flow. This means that such observer measures zero energy and momentum densities what corresponds to the Einstein equivalence principle. The same result was obtained in the FRLW  solution in TEGR \cite{EPT19,EPT_2020}.  Schwartzchild solution was considered in different coordinates: static and general radially freely falling coordinates. ``Turning off'' gravity in different coordinates we obtained different connections (different when expressed at the same coordinates). Such kind of ambiguities in ``turning off'' gravity principle was also found in TEGR \cite{EKPT_2021,EKPT_2021a}. However, in the Schwartzchild solution, in TEGR there were more physically expected results than in STEGR. In static case the situations in TEGR and STEGR are the same: Noether charge gives the correct black hole mass $M$  but non-zero Noether current for the freely falling observer {that} means that the Einstein equivalence principle is not satisfied. In the Lemaitre coordinates (general radially freely falling coordinates with $e=1$) it is impossible to ``turn off'' gravity in STEGR while in TEGR  the Einstein equivalence principle was obtained. In the general radially freely falling coordinates, in STEGR, the correct black hole mass $M$  was obtained, but not the Einstein equivalence principle for the freely falling observer; while in TEGR both the correct black hole mass $M$ and the Einstein equivalence principle for the freely falling observer were satisfied.

In a more complicated case of axially symmetric solutions in $f(T,B,\phi,X)$ theories our ``turning off gravity'' procedure elaborated in \cite{EPT19,EPT_2020} has been applied to
find the correct connection in \cite{Bahamonde_all_2021}. The resulting tetrads and connections solved the field equations (including antisymmetric parts of them) for Taub-NUT solution, but not for Kerr and C-metric solutions. It could be interesting to examine these solutions in constructing conserved quantities in the framework of STEGR with the ``turning off gravity'' procedure outlined in subsection \ref{turning_off}.

Can we satisfy both the correct black hole mass $M$ and the Einstein equivalence principle in STEGR? In fact, one can take more general set of flat torsionless connections without the  ``turning off'' gravity principle. Thus, the components of such a connection in general form were found in \cite{Hohmann:2019fvf}, and the most generic static spherical connections in symmetric
teleparallel geometry were worked out in \cite{Heisenberg_all_2021}. Basing on the results \cite{Hohmann:2019fvf,Heisenberg_all_2021} one can try to solve the equations of zero current for the  freely falling observer and condition for Noether's charge $M$ inside an unique set of appropriate teleparallel connections.

It is well known that, for example, cosmological and astrophysical models considered in GR, equivalently in STEGR, differ from those considered in $f(Q)$ theories \cite{Anagnostopoulos_2016}. Therefore, it would be interesting to apply the methods developed in STEGR to $f(Q)$ gravity and compare results of their applications. First, because $f(Q)$ theories are covariant ones we can apply the Noether procedure to derive the conservation laws and conserved quantities as well. At the moment, we do not know their connection with those in STEGR, and it would be interesting to study it. Second, the definitions of a flat torsionless connection in STEGR and in $f(Q)$ gravity differ. Recall the situation in TEGR and $f(T)$. The inertial spin connection in TEGR is defined by the external principle  of "turning off" gravity, without taking into account field equations, see \cite{EPT19,EPT_2020}. Unlike this, in covariant $f(T)$ theories (see, for example, \cite{Martin_2016}), inertial spin connection has to follow after solving the related system of field equations that can be not so simple. Concerning STEGR and $f(Q)$ theories, the situation has to be analogous. For example, it could be very difficult to find a related flat connection for the FRLW solution in the framework of $f(Q)$ theory. We think also that in the case of some spacetime symmetries in the process of the ``turning off’’ gravity one can satisfy the field equations in $f(Q)$ gravity, and it would be interesting to find such symmetries and then study the Noether conserved quantities for them.


We leave the above problems for a future work.



{\bf Acknowledgments.} AP has been supported by the Interdisciplinary Scientific and
Educational School of Moscow University “Fundamental and Applied Space Research”; EE
and AT are supported by RSF grant 21-12-00130. The authors are grateful to Martin  Kr\v s\v s\'ak and Laur J\"arv for helpful discussions and recommendations.
\bibliography{references}
\bibliographystyle{Style}

\end{document}